\documentclass[A4paper,twocolumn]{article}
\usepackage{graphicx}
\usepackage{amsmath}

\begin{document}

\title{Entanglement and genuine entanglement of three-qubit
Greenberger-Horne-Zeilinger diagonal states}
\author{Xiao-yu Chen$^1,$ Li-zhen Jiang$^1$, Ping Yu$^2$, Mingzhen Tian$^3$ \\
%EndAName
{\small {$^1$ College of Information and Electronic Engineering, Zhejiang
Gongshang University, Hangzhou, Zhejiang 310018, China }}\\
{\small {$^{2}$Department of Physics and Astronomy, University of Missouri,
Columbia, Missouri 65211, USA }}\\
{\small {$^3$School of Physics, Astronomy and Computational Sciences, George
Mason University, Fairfax, Virginia 22030, USA}}}
\date{}
\maketitle

\begin{abstract}
We analytically prove the necessary and sufficient criterion for
the full separability of three-qubit Greenberger-Horne-Zeilinger
(GHZ) diagonal states. The corresponding entanglement is exactly
calculable for some GHZ diagonal states and is tractable for the
others using the relative entropy of entanglement. We show that
the biseparable criterion and the genuine entanglement are
determined only by the biggest GHZ diagonal element regardless of
all the other smaller diagonal elements. We have completely solved
the entanglement problems of three-qubit GHZ diagonal states.

PACS number(s): 03.67.Mn; 03.65.Ud\newline
\end{abstract}

\textit{Introduction.---}Quantum entanglement, being a special form of
quantum superposition, possesses structures and properties intrinsically
different from any classical system. This difference depends on the number
of particles or states involved in the entanglement. For example, the test
of Bell inequality using two-qubit entangled states gives statistical
prediction, while the three-qubit Greenberger-Horne-Zeilinger (GHZ)
entangled states lead to a conflict with local realism non-statistically
\cite{Bell,GHZ}. Furthermore, the entangled states of more than two qubits
are more complicated because of a complex structure due to different ways
the qubits can be entangled \cite{Imoto}. Therefore new characteristic
methods for multi-qubit states are necessary to fully understand and
interpret their quantum behaviors in quantum mechanics.

Multipartite entanglement right now is at the core of quantum information,
and provides a critical resource for quantum secret sharing, quantum
error-correcting codes and quantum computation \cite{Horodecki,GuhneToth}. A
potential route to quantum computer is the manipulation of electrons trapped
in quantum-dot pairs \cite{Loss}. Recently, the first step revealing the
true scalability of spin-based quantum computing was taken by coherently
manipulating three individual electron spins confined in neighboring quantum
dots \cite{Gaudreau}, where the exchange interactions between the spins are
precisely controlled and entangled three-spin states are generated. To date,
multipartite entanglement has been observed in ion traps \cite{Blatt},
photon polarization \cite{Pan}, superconducting phase and circuit qubit
systems \cite{Martinis}, and nitrogen-vacancy centers in diamond \cite
{Neumann}. Due to detrimental decoherence effects and imperfections in
preparation, the multipartite entangled states prepared are usually mixed,
typically, so called GHZ diagonal states. With the experimental
realizations, a general separable criterion for GHZ mixed states is
important and desired.

One of the key issues is to determine whether the prepared states are
genuinely entangled or not entangled at all. Theoretical research has
concentrated on the separability and biseparability that characterize the
entanglement of the GHZ diagonal states \cite
{Dur,GuhneNJP,GuhneJM,Doherty,Eltschka,Kay,Guhne}. The criterion for
biseparability of three and four-qubit GHZ diagonal states have recently
been proved in the form of an inequality involving several density matrix
elements in the computational basis \cite{GuhneNJP}. We will show that this
criterion can be simplified to the largest GHZ diagonal component being
equal or less than 1/2. This is a significant step that leads to the
quantification of genuine entanglement. For the full separability of
three-qubit GHZ diagonal states, the sufficient criterion is proposed by
directly constructing the fully separable states \cite{Kay} and the
necessary criterion is proposed by the method of convex combinations \cite
{Guhne}. Numerical results strongly indicate that the two criteria should
coincide. However, it is inconclusive in the absence of an analytical proof.

In this letter, we will prove the unification of the criteria for full
separability. The full separability and biseparability provide a practical
and accurate method for complete classification of three-qubit GHZ diagonal
states, which now can be made into three categories: separable, entangled,
and genuinely entangled states discriminated by full separability and
biseparability, respectively. In addition to the complete entanglement
structure of three-qubit GHZ diagonal states, we also find a new group of
entangled states due to the cooperation of the off-diagonal elements in
computational basis. This cooperative entanglement may not be detected by
the commonly used positive partial transpose (PPT) criterion \cite{Peres}.
We will give the formula of genuine entanglement and reduce the
quantification of entanglement to algebraic calculations.

\textit{Necessary and sufficient condition of fully separable three-qubit
GHZ states}. ---The three-qubit GHZ diagonal states take the form
\begin{equation}
\rho =\sum_{k=1}^8p_k\left| GHZ_k\right\rangle \left\langle GHZ_k\right| ,
\label{wee0}
\end{equation}
where the $p_k$ form a probability distribution. The GHZ state basis
consists of eight vectors $\left| GHZ_k\right\rangle =\frac 1{\sqrt{2}%
}(\left| 0x_2x_3\right\rangle \pm \left| 1\overline{x}_2\overline{x}%
_3\right\rangle ),$ with $x_i,\overline{x}_i\in \{0,1\}$ and $x_i\neq
\overline{x}_i.$ In the binary notation, $k-1=0x_2x_3\ $for the '+' states
and $k-1=1\overline{x}_2\overline{x}_3$ for the '-' states. Using Pauli
matrices $X,Y,Z$ and $2\times 2$ identity matrix $I,$ the GHZ diagonal
states can be written as
\begin{eqnarray}
\rho &=&\frac 18[III+\lambda _2ZZI+\lambda _3ZIZ+\lambda _4IZZ+\lambda _5XXX
\nonumber \\
&&+\lambda _6YYX+\lambda _7YXY+\lambda _8XYY],  \label{wee1}
\end{eqnarray}
where tensor product symbols are omitted.

The sufficient condition of full separability of $\rho $\ is \cite{Guhne}
\begin{equation}
1-\left| \lambda _{-}\right| -\mu \geq 0,  \label{wee2}
\end{equation}
or
\begin{equation}
1-\left| \lambda _{-}\right| -\left| \lambda _5\right| -\left| \lambda
_6\right| -\left| \lambda _7\right| -\left| \lambda _8\right| \geq 0
\label{wee3}
\end{equation}
where $\lambda _{-}=\min \{\lambda _2+\lambda _3+\lambda _4,$ $\lambda
_2-\lambda _3-\lambda _4,$ $-\lambda _2+\lambda _3-\lambda _4,$ $-\lambda
_2-\lambda _3+\lambda _4,\},$ and
\begin{equation}
\mu =\frac{\sqrt{(\lambda _5\lambda _6+\lambda _7\lambda _8)(\lambda
_5\lambda _7+\lambda _6\lambda _8)(\lambda _5\lambda _8+\lambda _6\lambda _7)%
}}{\sqrt{\lambda _5\lambda _6\lambda _7\lambda _8}}.  \label{wee4}
\end{equation}
The sufficient condition comes from an explicit construction of $\rho $ in a
fully separable manner \cite{Kay} \cite{Guhne}.

Let us consider the $8\times 8$ density matrix $\rho $ with entries $\rho
_{ij}$ in the basis $\left| 000\right\rangle ,\left| 001\right\rangle
,\ldots ,\left| 111\right\rangle $ which are ordered in the canonical way.
For GHZ diagonal states, the only nonzero elements of $\rho $ are $\rho
_{ii} $ and $\rho _{i,9-i}$ $(i=1,\ldots ,8),$ and further we have $\rho
_{ii}$ $=$ $\rho _{9-i,9-i}.$

The necessary condition for the full separability of GHZ diagonal
state $\rho $\ \cite{Guhne}\ can be written as
\begin{equation}
\left| \mathcal{L}(\rho ,\overrightarrow{X})\right| \leq C(\overrightarrow{X}%
)\kappa  \label{wee5}
\end{equation}
where $\kappa =\min \{\rho _{ii}$ $\left( 1\leq i\leq 4\right) \},%
\overrightarrow{X}=(\delta ,\alpha ,\beta ,\gamma )\ $is a real vector, $%
\mathcal{L}(\rho ,\overrightarrow{X})=\delta \rho _{18}+\alpha \rho
_{27}+\beta \rho _{36}+\gamma \rho _{54},$ and
\begin{eqnarray}
C(\overrightarrow{X}) &=&\sup_{a,b,c}[\delta \cos (a+b+c)+\alpha \cos
(a)+\beta \cos (b)  \nonumber \\
&&+\gamma \cos (c)].  \label{wee6}
\end{eqnarray}
Here $a,b$,and $c$ are the angles. The relationship among the density matrix
entries $\rho _{ij}$ and the parameters $\lambda _k$ of a GHZ diagonal state
is a simple linear transformation:
\begin{eqnarray}
(\lambda _5,-\lambda _6,-\lambda _7,-\lambda _8) &=&4(\rho _{18},\rho
_{36},\rho _{27},\rho _{54})H_2,  \label{rtt1} \\
\kappa &=&(1-\left| \lambda _{-}\right| )/8,  \label{rtt2}
\end{eqnarray}
where $H_2$ is the $4\times 4$ Hadamard matrix. Consider the trivial case
that $\overrightarrow{X}$ is a positive vector, $C(\overrightarrow{X})\ $%
achieves its maximum $\delta +\alpha +\beta +\gamma $ when $a=b=c=0.$ If the
off-diagonal elements of $\rho $ are all positive, the left-hand side of
inequality (\ref{wee5}) is the probability mixture of the off-diagonal
elements. If two of the off-diagonal elements $\rho _{18},\rho _{27},\rho
_{36},\rho _{54}$ are negative, the state can be transformed to a state with
all positive off-diagonal elements by local operations and classical
communication (LOCC). For these cases, the inequality (\ref{wee5}) can be
rewritten as
\begin{equation}
\max \{\left| \rho _{i,9-i}\right| ,\left( 1\leq i\leq 4\right) \}\leq \min
\{\rho _{ii},\left( 1\leq i\leq 4\right) \}.  \label{wee7}
\end{equation}
It is just the PPT criterion. In the case of even number of negative $\rho
_{18},\rho _{27},\rho _{36},\rho _{54}$, namely $\prod_{i=5}^8\lambda _i\leq
0,$ PPT criterion is necessary and sufficient for the full separability \cite
{Kay}.

What left is to show that if the inequality (\ref{wee2}) can be derived from
(\ref{wee5}) or vice versa when $\prod_{i=5}^8\lambda _i>0$. When one of the
components of $\overrightarrow{X}$ is negative (for definiteness, one may
choose $\gamma <0$), the solutions have been obtained \cite{Guhne} for the
angles $a,b,c$ to optimize $C(\overrightarrow{X})$ for any given
coefficients $\delta ,\alpha ,\beta ,\gamma $. The solutions fulfill the
following equations derived from (\ref{wee6})
\begin{equation}
\delta \sin d=-\alpha \sin a=-\beta \sin b=\left| \gamma \right| \sin c,
\label{wee8}
\end{equation}
where $d\equiv a+b+c.$ Without loss of generality, we consider the case
where $\rho _{1,8},\rho _{3,6},\rho _{2,7}$ are non-negative (this is always
possible by LOCC), denote $(x,y,z)=(\delta ,\alpha ,\beta )/\left| \gamma
\right| ,$ and define
\begin{eqnarray}
f(x,y,z) &\equiv &\frac{\left| \mathcal{L}(\rho ,\overrightarrow{X})\right|
}{C(\overrightarrow{X})\kappa }  \nonumber \\
&=&\frac{(x\rho _{18}+y\rho _{36}+z\rho _{27}-\rho _{54})/\kappa }{x\cos
d+y\cos a+z\cos b-\cos c}.  \label{wee9}
\end{eqnarray}
For the outmost surface of the full separability set determined by the
necessary condition, we have the following equations
\begin{eqnarray}
\frac{\partial f(x,y,z)}{\partial x} &=&\frac{\partial f(x,y,z)}{\partial y}=%
\frac{\partial f(x,y,z)}{\partial z}=0,  \label{wee10} \\
f(x,y,z) &=&1.  \label{wee11}
\end{eqnarray}
The solution to equations (\ref{wee10}) is
\begin{equation}
\rho _{18}=\kappa \cos d,\rho _{36}=\kappa \cos a,\rho _{27}=\kappa \cos
b,\rho _{54}=\cos c,  \label{wee15}
\end{equation}
where we have used equations (\ref{wee8}), (\ref{wee11}) and the fact that $%
d=a+b+c$ and $a,b,c,d$ are functions of $x,y,z$ in equation (\ref{wee10}).

Substituting the solution (\ref{wee15}) into equation (\ref{rtt1}) to obtain
$\lambda _i$ $(i=5,...,8)$, and using of (\ref{wee4}) and (\ref{rtt2}), we
have
\begin{equation}
\mu =1-\left| \lambda _{-}\right| ,  \label{wee16}
\end{equation}
Hence the outmost surface of fully separable state set determined by the
necessary condition takes the form of Eq. (\ref{wee16}), which is also the
outmost surface of fully separable state set determine by the sufficient
condition (\ref{wee2}). In either odd or even number of negative parameters $%
\delta ,\alpha ,\beta ,\gamma ,$ it is shown that the necessary criterion of
full separability is also sufficient. Thus we obtain the necessary and
sufficient criterion for fully separable of three-qubit GHZ diagonal states.

Consider the case of\textit{\ }$\rho _{18}=\rho _{36}=\rho _{27},$ we have
the simple solution of $b=a,$ $c=-3a,$ $d=-a$ $(a\in [0,\pi /3])$ for the
border curve of the full separability. More explicitly, we have the equation
of border curve
\begin{equation}
\frac{\rho _{54}}\kappa =4(\frac{\rho _{18}}\kappa )^3-3\frac{\rho _{18}}%
\kappa .  \label{wee17}
\end{equation}
The border curve is shown in Fig.1.

\begin{figure}[tbp]
\includegraphics[ trim=0.000000in 0.000000in -0.138042in 0.000000in,
height=2.5in, width=3.5in ]{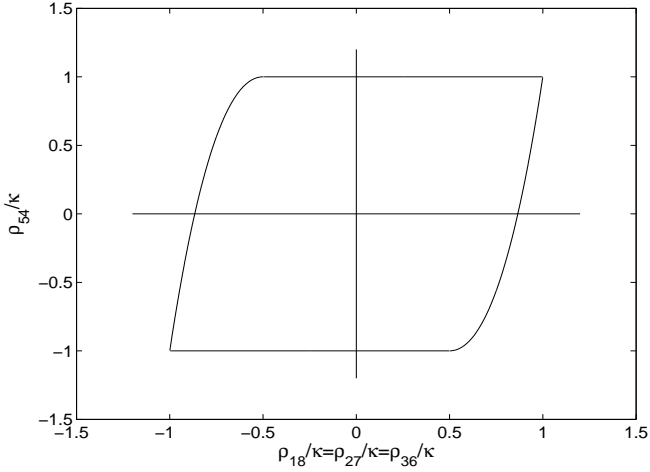}
\caption{The border curve of full separability}
\end{figure}

\textit{Entanglement:---}Entanglement of a given entangled state $\varrho $
can be quantified by its relative entropy of entanglement (REE) as, $%
E=\min_{\sigma \in SEP}S(\varrho \left\| \sigma \right. ),$ here $SEP$ is
the fully separable state set. The relative entropy is defined as, $%
S(\varrho \left\| \sigma \right. )=:Tr\varrho \log _2\varrho -Tr\varrho \log
_2\sigma .$ The separable state achieves the minimal relative entropy for
state is called its closest state. Follow the same reason given in ref \cite
{Vedral}, the closest state for a three-qubit GHZ diagonal state $\rho $
must also be a GHZ diagonal state in the form of $\sigma
=\sum_{k=1}^8q_k\left| GHZ_k\right\rangle \left\langle GHZ_k\right| $, where
the $q_k$ form a probability distribution. As the result, the REE can be
calculated as,
\begin{equation}
E=\sum_{k=1}^8p_k\log _2(\frac{p_k}{q_k}).  \label{rtt3}
\end{equation}
The closest state $\sigma $ is at the boundary of the fully separable states
due to the following reasoning: Suppose $\sigma ^{\prime }$ is an inner
state within the fully separable state set. The entangled state $\rho $ is
at the outside. A link between $\sigma ^{\prime }$ and $\rho $ should
intersect with the boundary of the fully separable state set. We have a
border state $\sigma =\lambda \sigma ^{\prime }+(1-\lambda )\rho $ for some $%
\lambda \in (0,1).$ Then $S(\rho \left\| \sigma \right. )<$ $\lambda S(\rho
\left\| \sigma ^{\prime }\right. )+(1-\lambda )$ $S(\rho \left\| \rho
\right. )=$ $\lambda S(\rho \left\| \sigma ^{\prime }\right. )<$ $S(\rho
\left\| \sigma ^{\prime }\right. ).$ The first inequality comes from the
convexity of negative logarithmic function. Hence an inner separable state
is not a closest state. The parameters of a closest state can be written as $%
q_k=s_k+\kappa _c\cos \theta _k$ and $q_{9-k}=s_k-\kappa _c\cos \theta _k$
for $k=1,\cdots 4,$ where $s_k\geq 0,$ $\kappa _c=\min_k\{s_k\},$ $\mathbf{%
\theta }=(d,a,b,c).$ The free variables are $a,b,c$ and three of $s_k$. We
have six free variables and six equations derived from the extremal of (\ref
{rtt3}). REE is tractable. In some special cases, analytical solutions can
be obtained as we will show below.

If an entangled GHZ diagonal state $\rho $ has the symmetry of $p_1=p_2=p_3\
$and $p_6=p_7=p_8$, we will prove that the closest state should have the
same symmetry of $q_1=q_2=q_3\ $and $q_6=q_7=q_8.$ Suppose that a
nonsymmetric state $\sigma _1=\sum_{k=1}^8q_k\left| GHZ_k\right\rangle
\left\langle GHZ_k\right| $ is the closest state of $\rho $. By cycling $%
q_1,q_2,q_3$ and $q_6,q_7,q_8,$ we obtain $\sigma _2$ and $\sigma _3$ which
also are closest states of $\rho .$ Let $\overline{\sigma }=\sum_{k=1}^8%
\overline{q}_k\left| GHZ_k\right\rangle \left\langle GHZ_k\right| =\frac
13(\sigma _1+\sigma _2+\sigma _3),$ with $\overline{q}_1=\overline{q}_2=%
\overline{q}_3=$ $\frac 13(q_1+q_2+q_3),$ $\overline{q}_6=\overline{q}_7=%
\overline{q}_8=$ $\frac 13(q_6+q_7+q_8)$, then $\overline{\sigma }$ is fully
separable since it is a probability mixture of fully separable states. Note
that $\frac 13(q_1+q_2+q_3)>\sqrt[3]{q_1q_2q_3}$. We have $S(\rho \left\|
\overline{\sigma }\right. )<S(\rho \left\| \sigma _1\right. ),$ that is a
contradiction. The symmetric closest state $\sigma $ is with $\sigma
_{11}=\sigma _{22}=\sigma _{33}$ and $(\sigma _{18},\sigma _{36},\sigma
_{27},\sigma _{54})$ $=\kappa _c(\cos \theta ,\cos \theta ,\cos \theta ,\cos
3\theta ),$ where $\theta \in [0,\frac \pi 3]$, and $\kappa _c=\min \{\sigma
_{11},\sigma _{44}\}$. Let $\sigma _{11}=\frac 18+\frac 13\xi ,$ $\sigma
_{44}=\frac 18-\xi ,$ where $\xi \in [-\frac 38,\frac 18]$ $,$ we obtain the
two optimal equations with variables $\theta $, $\xi .$ The equations are
not analytically solvable in general. However, if the closest state is with $%
\kappa _c=\frac 18,$ exact solution can be obtained. The derivative of (\ref
{rtt3}) on $\theta $ leads to
\begin{equation}
(\frac 14-\rho _{11})t^3-\rho _{18}t^2-(\frac 34-4\rho _{11})t+\rho
_{18}-\rho _{54}=0,  \label{wee19}
\end{equation}
where $t=2\cos \theta $. We then properly choose $\theta $ from the three
solutions of (\ref{wee19}). Note that the derivative of $\xi $ may not exist
at $\xi =0,$ we have to calculate the left and right derivatives, which are
\begin{eqnarray*}
\frac{\partial S(\rho \left\| \sigma \right. )}{\partial \xi }\left| _{\xi
=0^{-}}\right. &=&c_1(\theta )[\cos 3\theta (\cos 3\theta -8\rho
_{54})+8\rho _{44}-1], \\
\frac{\partial S(\rho \left\| \sigma \right. )}{\partial \xi }\left| _{\xi
=0^{+}}\right. &=&c_2(\theta )[\cos \theta (8\rho _{18}-\cos \theta
)+1-8\rho _{11}],
\end{eqnarray*}
where we have used equation (\ref{wee19}) to simplify the expressions. $%
c_1(\theta )=\frac 8{3\sin ^23\theta }$ and $c_2(\theta )=\frac{8(4\cos
^2\theta -1)}{\sin ^23\theta }$ are positive factors for $\theta \in
[0,\frac \pi 3]$.

Analytical solutions exist for the states with $\rho _{11}=\rho _{44}=\frac
18$ when $\theta \in (\frac \pi 6,\frac \pi 3].$ The entanglement can be
obtained exactly. We verify that $\cos \theta <8\rho _{18},\cos 3\theta
>8\rho _{54}$ are true for all the entangled states. The left derivative is
negative since $\cos 3\theta <0$ for $\theta \in (\frac \pi 6,\frac \pi 3]$
and the right derivative is positive. When $\rho _{54}=-\rho _{18}$, the
solution of (\ref{wee19}) is $\theta =\frac \pi 4.$ The REE is
\[
E=\frac{1+8\rho _{18}}2\log _2\frac{1+8\rho _{18}}{1+\frac 1{\sqrt{2}}}+%
\frac{1-8\rho _{18}}2\log _2\frac{1-8\rho _{18}}{1-\frac 1{\sqrt{2}}}.
\]

Analytical solutions also exist for states with $\rho _{18}=\rho _{11},$ $%
\rho _{54}=-\rho _{44}.$ Then $p_1=2\rho _{11},$ $p_5=2\rho _{44},$ $%
p_4=p_8=0.$ We have four kinds of candidate closest states: (i) $\kappa
_c=\frac 18,$ (ii) $\kappa _c=\sigma _{1,1}<\frac 18,$(iii) $\kappa
_c=\sigma _{4,4}<\frac 18,$ (iv) PPT boundary state. Here we consider $%
p_5>p_1$ (The case $p_5<p_1$ can be solved similarly). For the first closest
state candidate, the relative entropy is simply
\begin{equation}
S(\rho \left\| \sigma \right. )=3p_1\log _2\frac{p_1}{1+\cos \theta }%
+p_5\log _2\frac{p_5}{1-\cos 3\theta }.  \label{wee24}
\end{equation}
From $\frac{dS(\rho \left\| \sigma \right. )}{d\theta }=0$, we have $(2\cos
^2\theta +\Delta \cos \theta -1)(2\cos \theta +1)=0,$where $\Delta =\frac{%
p_5-p_1}{p_5+p_1}.$ The solution is $\cos \theta =\frac 14(\sqrt{8+\Delta ^2}%
-\Delta ).$ To check that the closest state locates at $\kappa _c=\frac 18$,
we prove that the right derivative is always positive and the left
derivative is negative only when $p_1>p_0=\frac 1{12}(3+\cos 3\theta
)\approx $ $0.1718.$ Hence candidate (i) is the solution when $p_1>p_0$. For
$p_1\leq p_0$, we consider candidate (ii). We have $S(\rho \left\| \sigma
\right. )=3p_1\log _2\frac{p_1}{\kappa _c(1+\cos \theta )}+p_5\log _2\frac{%
p_5}{\frac 12-\kappa _c(3+\cos 3\theta )}$. Then $\frac{\partial S(\rho
\left\| \sigma \right. )}{\partial \kappa _c}=0$ gives $\kappa _c=\frac{3p_1%
}{2(3+\cos 3\theta )}$ .$\ $ We have
\begin{equation}
S(\rho \left\| \sigma \right. )=1+3p_1\log _2\frac{(3+\cos 3\theta )}{%
3(1+\cos \theta )}.  \label{wee25}
\end{equation}
The optimal equation is $4\cos ^3\theta +6\cos ^2\theta -3=0$, the solution
is $\cos \theta =\frac 12(\sqrt[3]{2+\sqrt{3}}+\sqrt[3]{2-\sqrt{3}}%
-1)\approx 0.5979$. The condition $\kappa _c<\frac 18\ $is equivalent to $%
p_1<p_0$. The candidate (iii) does not give rise to a further small value of
the relative entropy because $\frac{\partial S(\rho \left\| \sigma \right. )%
}{\partial \kappa _c}<0$ for all $\theta \in [0,\frac \pi 3].$ The candidate
(iv) can also be removed.

\textit{Genuine entanglement}:---The necessary and sufficient criterion has
been proven for biseparability of a three-qubit GHZ diagonal state \cite
{GuhneNJP}. It has the form of
\begin{equation}
\left| \rho _{18}\right| \leq \sqrt{\rho _{22}\rho _{77}}+\sqrt{\rho
_{33}\rho _{66}}+\sqrt{\rho _{44}\rho _{55}}.  \label{wee28}
\end{equation}
The criterion can be rewritten as
\begin{equation}
\max \{p_i\}\leq \frac 12,  \label{wee29}
\end{equation}
since for three-qubit GHZ diagonal state, we have $\rho _{ii}=\rho
_{9-i,9-i},$ and $p_1=\rho _{1,1}+\left| \rho _{1,8}\right| .$ When $%
p_1>\frac 12$, it is followed from the reasoning in the entanglement of the
Bell diagonal state \cite{Vedral} that the state is not biseparable. The
genuine entanglement measured by REE is $E_{genuine}(\rho )=\min_{\sigma \in
BISEP}S(\rho \left\| \sigma \right. ),$ where $BISEP$ is the biseparable
state set. Hence
\begin{equation}
E_{genuine}(\rho )=1+p_1\log _2p_1+(1-p_1)\log _2(1-p_1).  \label{wee30}
\end{equation}
Also criterion (\ref{wee29}) and Eq. (\ref{wee30}) are true for four qubit
GHZ state as the necessary and sufficient criterion has been proven \cite
{GuhneNJP}. The mixture of N-qubit GHZ state and white noise is $\rho
^{(ghzN)}=(1-p)\left| GHZN\right\rangle $ $\left\langle GHZN\right| +p%
\mathbf{1}/2^N$. The state is genuinely entangled iff $0\leq
p<1/[2(1-2^{-N})].$ The probability of $\left| GHZN\right\rangle $ component
is $p_1=(1-p)+p/2^N=1-p(1-2^{-N})>\frac 12.$ The biseparable (yet
inseparable under bipartitions) condition can also be written in the form of
inequality (\ref{wee29}). Therefore its genuine REE can be expressed as Eq.(%
\ref{wee30}).

\textit{Summary:---}For three-qubit GHZ diagonal state, the fully separable
criterion has been strictly proven to be necessary and sufficient. We find
the exact boundary states for the fully separable state set. The free
variable solution of boundary states make the calculation of the relative
entropy of entanglement easy. The relative entropy of entanglement is
exactly calculated for the symmetric states $\rho $ ( $%
p_1=p_2=p_3,p_6=p_7=p_8$) of type (i) $\rho _{ii}=\frac 18$ ($i=1,\cdots 8$)
(diagonal elements are equal in computational basis) and type (ii) $%
p_4=p_8=0 $. The closest states $\sigma $ can be either with
$\sigma _{ii}=\frac 18$ ($i=1,\cdots 8$) or not for both type (i)
and type (ii) entangled states. There are many subtleties in
obtaining the closest states due to the possible non-existence of
the derivative of the relative entropy. We also give the genuine
entanglement of GHZ diagonal states in terms of the relative
entropy of entanglement. The genuine entanglement is determined by
the biggest GHZ diagonal component only. The genuine entanglement
formula obtained is easily extended to $N$ particle GHZ diagonal
states. The fully separable criterion can be applied as a
necessary criterion for the separability of any three-qubit state
by filtering it to a GHZ diagonal state. Further works on the
separability and entanglement of more than three qubit GHZ
diagonal states are desirable.

XYC and LZJ thank the support of the National Natural Science Foundation of
China (Grant No. 60972071).

\end{document}